\documentstyle[11pt,epsf,newpasp,twoside]{article} 
\newcommand{\kms}{km\,s$^{-1}$}

\begin{document}
\title{Double degenerates and progenitors of supernovae type Ia}

%
\author{R.~Napiwotzki}
\affil{Dept.\ of Physics \& Astronomy, University of Leicester,
  Leicester, UK}
\author{L.~Yungelson}
\affil{Inst.\ of Astronomy of the Russian Academy of Sciences, 
  Moscow, Russia}
\author{G.~Nelemans}
\affil{Institute of Astronomy, Cambridge, UK}
\author{T.R.~Marsh}
\affil{Department of Physics, University of Warwick, Coventry, UK}
\author{B.~Leibundgut, A.~Renzini}
\affil{European Southern Observatory, Garching, Germany}
\author{D.~Homeier}
\affil{Dept.\ of Physics \& Astronomy,
        University of Georgia, Athens, GA, USA}
\author{D.~Koester, S.~Moehler}
\affil{Inst.\ f\"ur Theo.\ Physik und Astrophysik, 
  Universit\"at Kiel, Kiel, Germany}
\author{N.~Christlieb, D.~Reimers}
\affil{Hamburger Sternwarte, Universit\"at Hamburg, 
  Hamburg, Germany}
\author{H.~Drechsel, U.~Heber, C.~Karl, E.-M.~Pauli}
\affil{Remeis-Sternwarte, Universit\"at Erlangen-N\"urnberg,
  Bamberg, Germany}

\begin{abstract}
We report on systematic radial velocity surveys for white dwarf --
white dwarf binaries (double degenerates -- DDs) including SPY (ESO 
Supernovae Ia progenitor survey) recently carried out at the VLT.
A large sample of DD will allow us to put strong constrains on the
phases of close binary evolution of the progenitor systems and to
perform an observational test of the DD scenario for supernovae of
type Ia. We explain how parameters of the binaries can be derived from various methods. Results for a sample of DDs are presented and discussed.
\end{abstract}

\section{Introduction}

Supernovae of type Ia (SN\,Ia) play an outstanding role for our
understanding of galactic evolution and the determination of the
extragalactic distance scale.  However, the nature of their
progenitors is not yet settled (e.g.\ Livio 2000). SN\,Ia are observed
in all types of galaxies, including elliptical galaxies containing
only old stellar populations. The light curves of SN\,Ia are dominated
by the decay of radioactive material synthesised in the explosion
(mainly nickel, decaying to iron). The rapid evolution of SN\,Ia light
curves indicates that the precursors of these supernovae must be
compact objects of small mass with little mass holding back the
gamma-rays produced by the radioactive decay. The only candidate,
which can fulfil these observational constraints, is the thermonuclear
explosion of a white dwarf (WD).

According to the current consensus this happens when the WD grows to
the Chandrasekhar mass of $\approx$$1.4M_\odot$. Since no way is known
how this can happen to a single WD, this can only be achieved by mass
transfer in a binary system. Several channels have been identified as
possibly yielding such a critical mass. They can be broadly grouped
into two classes. The single degenerate (SD) channel (Whelan \& Iben
1973) in which the WD is accompanied by either a main sequence star, a
(super)giant, or a helium star, as mass donor and the double
degenerate (DD) channel where the companion is another WD (Webbink 1984;
Iben \& Tutukov 1984).
Close DDs radiate gravitational waves, which results in a shrinking
orbit due to the loss of energy and angular momentum. If the initial
separation is close enough (orbital periods below $\approx$10\,h), a DD system
could merge within a Hubble time, and if the combined mass exceeds the
Chandrasekhar limit the DD would qualify as a potential SN\,Ia
progenitor. 

\section{Surveys for close DD}

The DD scenario for the progenitors was proposed many years ago. So
far, no SN\,Ia progenitor has been identified, which is not really
surprising considering the rareness of SNe\,Ia. The orbital velocity
of WDs in potential SN\,Ia progenitor systems must be large
($>150$\,km/s) making radial velocity (RV) surveys of WDs the
most promising detection method.
Most WDs are of the hydrogen-rich spectral type DA,
displaying broad hydrogen Balmer lines. The remaining WDs are
of non-DA spectral types (e.g.\ DB and DO) and
their atmospheres contain no or very little hydrogen.
Accurate RV measurements are possible for DA WDs
thanks to sharp cores of the H$\alpha$ profiles caused by NLTE
effects.

The first systematic search for DDs among white dwarfs was performed
by Robinson \& Shafter (1987). They applied a photometric technique
with narrow band filters centred on the wings of H$\gamma$ or He\,I
4471\AA\ for DA and DB WDs, respectively. RV velocity variations
should produce brightness variations in these filters.  This survey
investigated 44 WDs, but no RV variable systems were detected.  A few
years later Bragaglia et al.\ (1990) and Foss, Wade, \& Green (1991)
carried out spectroscopic investigations with moderate resolution and
signal-to-noise. While Foss et al.\ observed 25 WDs without detecting
an RV variable DD, Bragaglia checked 54 stars with one definitive
detection and four more uncertain candidates of which two were
confirmed by later observations. However, one of the confirmed
candidate was later reclassified as subdwarf~B star. The low number of
detections prompted Bragaglia et al.\ (1990) to state that DDs, at least
those with DA components, are unlikely precursors of SN\,Ia.  Typical
accuracies for these three investigations were $40\ldots 50$\,km/s,
which is only good enough to detect systems with periods of
$\approx$12\,h (if inclination angle or phase differences are not too
unfavourable), but not for longer period systems

After these frustrating results Marsh, Dhillon, \& Duck (1995) chose a
different approach. They selected seven low mass WDs, which
have a He core instead of the common C/O core, for their observations.
Since our Universe is too young for the formation of He core WDs
by single star evolution, it is expected that these low mass
WDs reside in binaries. Indeed, Marsh et al.\ detected
orbital RV variations in five of the seven systems. This result
dramatically increased the number of then known DDs and is consistent
with all He WDs residing in close binaries. Two larger
surveys were performed in the late nineties: one by Saffer, Livio, \&
Yungelson (1998) and one by Maxted \& Marsh (1999) and Maxted, Marsh,
\& Moran (2000a).  Saffer et al.\ surveyed 107 WDs with a
modest accuracy of $25$\,km/s and detected 18 candidates in two
quality classes. Some were not confirmed in later studies.  The
combined sample of Maxted \& Marsh (1999) and Maxted et al.\ (2000a)
contains 117 WDs with quite good RV accuracy of $2\ldots
3$\,km/s. However, in spite of their high accuracy they detected only
four good candidates, a detection rate much lower than in the Saffer
et al.\ sample.

Combining all the surveys about 200 WDs were checked for RV
variations with sufficient accuracy yielding 18 DDs with periods
$P<6.3$\,d (see Marsh 2000 for a compilation). However, none of these
systems seems massive enough to qualify as a SN\,Ia precursor. This is
not surprising, as theoretical simulations suggests that only a few
percent of all DDs are potential SN\,Ia progenitors (Iben, Tutukov, \&
Yungelson 1997; Nelemans et al.\ 2001). Note that some of the surveys
were even biased against finding SN\,Ia progenitors, because they
focused on low mass WDs. It is obvious that larger
samples are needed for statistically significant tests.

Recently, subdwarf B (sdB) stars with WD components have been
proposed as potential SNe\,Ia progenitors by Maxted et al.\ (2000b),
who announced the serendipitous discovery of a massive WD companion
of the sdB KPD\,1930+2752. If the canonical sdB mass of $0.5M_\odot$ is
adopted, the mass function the mass function yields a minimum 
total mass of the system in excess of the Chandrasekhar
limit. 
Since this system will merge in less than a Hubble time, this
makes KPD\,1930+2752 a SN\,Ia progenitor candidate (although this
interpretation has been questioned by Ergma et al.\ 2001).

\section{The SPY project}

The surveys mentioned above were performed with $3\ldots 4$\,m class
telescopes. A significant extension of the sample size without the use
of larger telescopes would be difficult due to the limited number of
bright WDs. This situation changed after the ESO VLT became
available. 
In order to perform a definitive test of the DD scenario we 
embarked on a large spectroscopic survey  of $\approx$1000 WDs 
(ESO {\bf S}N \,Ia
{\bf P}rogenitor surve{\bf Y} -- SPY). 
SPY has overcome the main limitation of all efforts so far to detect
DDs that are plausible SN~Ia precursors: the samples of surveyed
objects were too small.  

Spectra were taken with the high-resolution 
UV-Visual Echelle Spectrograph (UVES) of
the UT2 telescope (Kueyen) of the ESO VLT in service mode. 
Our instrument setup 
provided nearly complete spectral coverage from 3200\,\AA\ to 
6650\,\AA\
with a resolution $R=18500$ (0.36\,\AA\ at H$\alpha$). 
Due to the nature of
the project, two spectra at different, ``random'' epochs separated 
by at least one day were observed.
We routinely measure RVs with an accuracy of $\approx 2$\,\kms\
or better, therefore running only a very small risk of missing a merger
precursor, which have orbital velocities of 150\,\kms\ or higher.  
A detailed description of the SPY project can be found in Napiwotzki
et al.\ (2001a).

\begin{figure}[tbh]
\plottwo{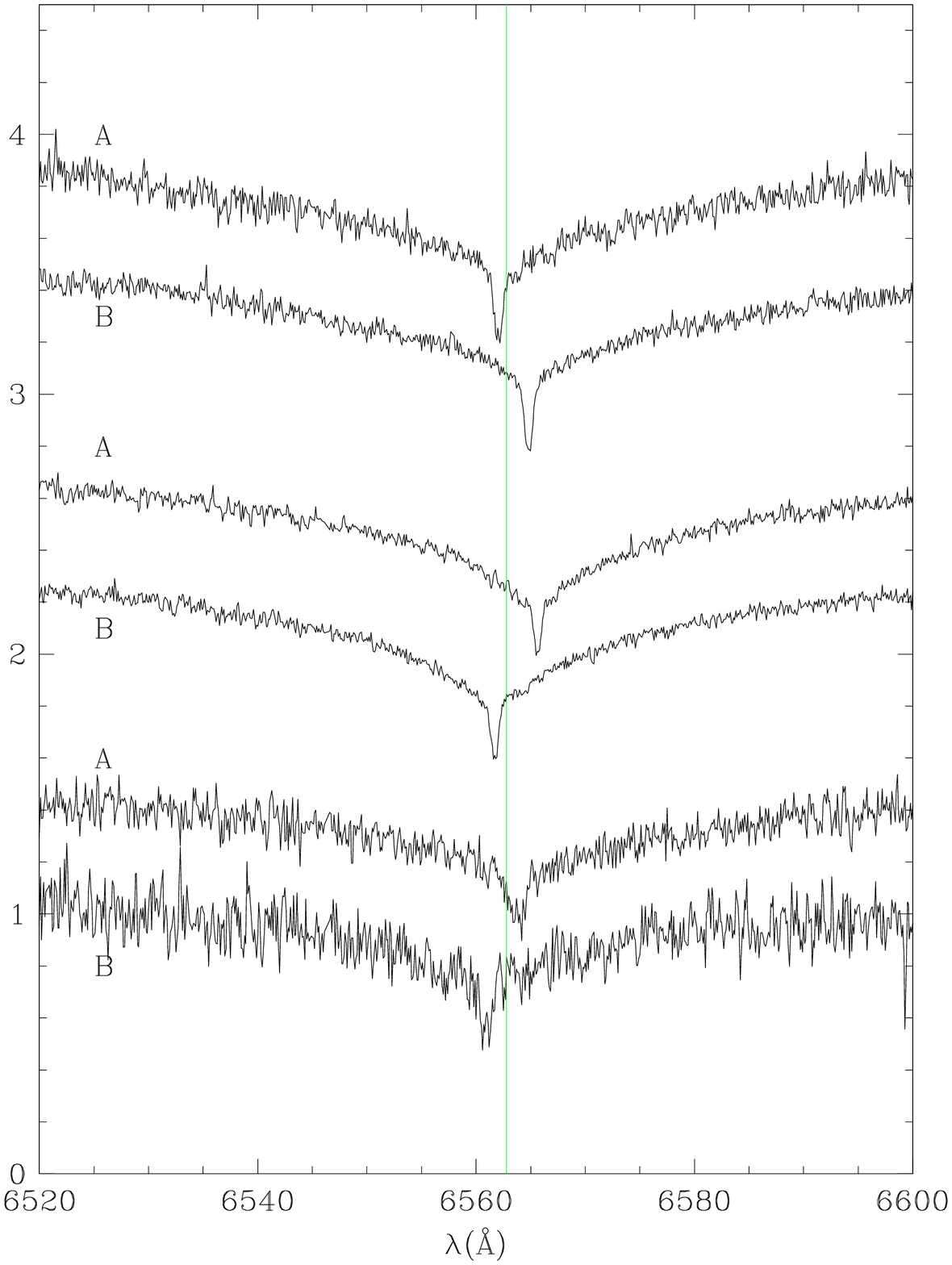}{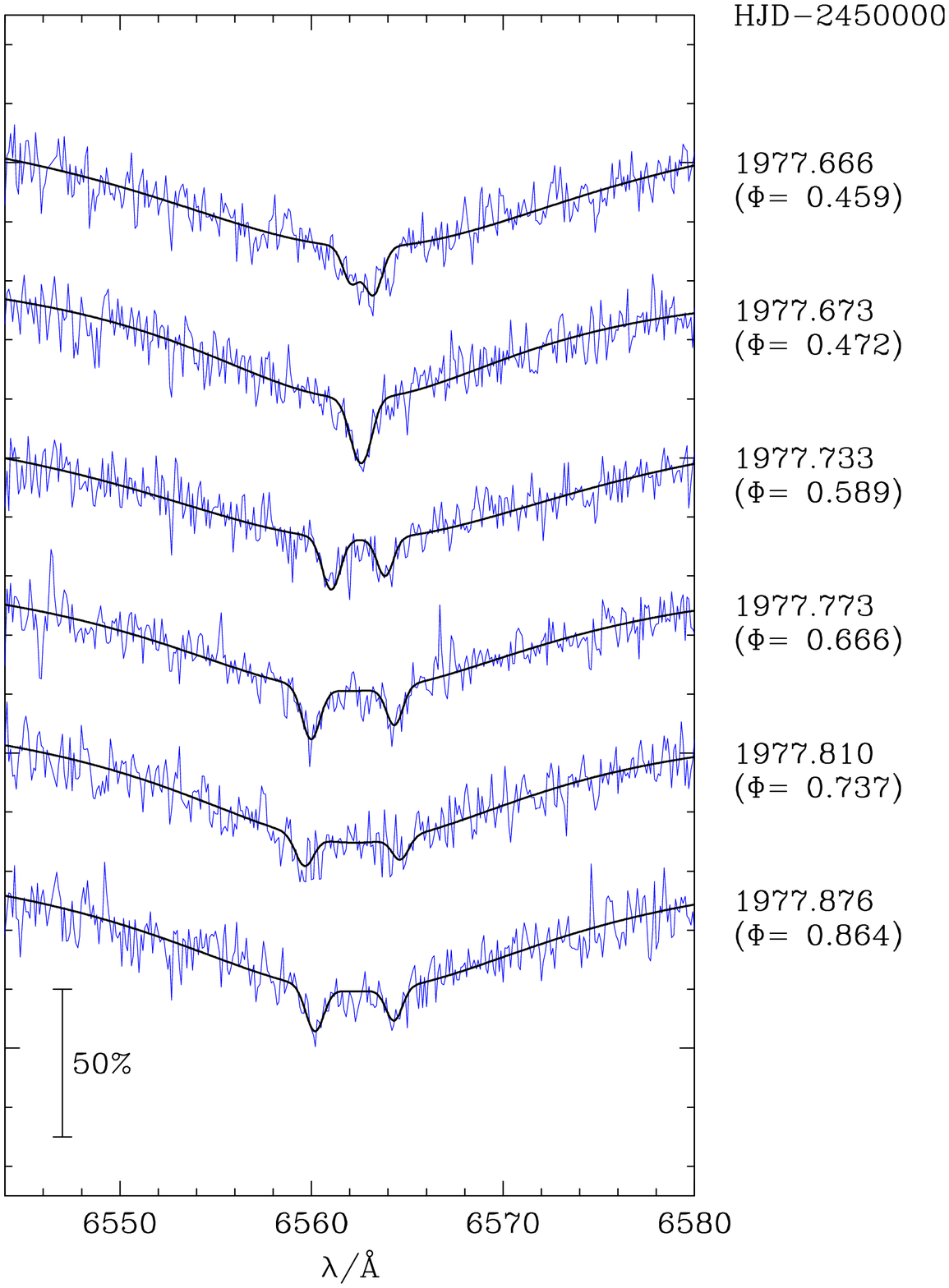}
\caption{{\it Left:} Three single-lined RV variable DDs from
our VLT survey. The green line marks the rest wavelength of H$\alpha$.
{\it Right:} H$\alpha$ spectra of HE\,1414-0848 
covering 5 hours during one night
together with a fit of the line cores. The numbers indicate the Julian
date of the exposures and the orbital phase $\phi$.
The spectra are slightly rebinned (0.1\,\AA) without degrading the
resolution. }
\label{f:HE1414ha}
\label{f:halpha}
\end{figure}

\begin{figure}[bt]
\plotone{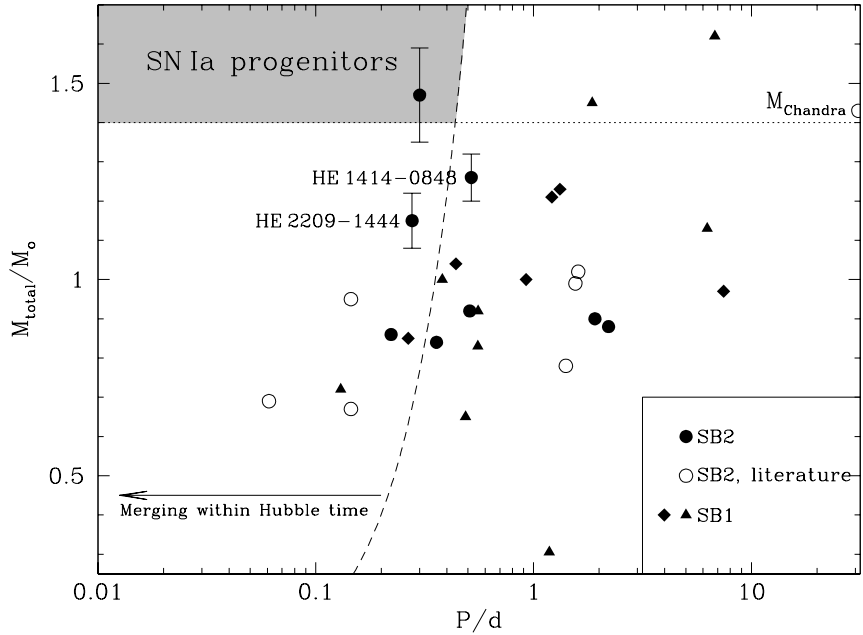}
\caption{Periods ($P$) and 
system masses ($M_{\mathrm{total}}$) determined from follow-up observations of
DDs from SPY. Results for double-lined systems 
are compared to previously known systems. 
The other DD systems are single-lined (triangles: WD primaries; diamonds:
sdB primaries). The masses of the unseen companions are
estimated from the mass function for the expected average inclination angle 
($i=52^\circ$).}
\label{f:pm}
\end{figure}

The large programme has finished at the end of March 2003. A total of
1014 stars were observed. This corresponds to 75\% of the known WDs
accessible by VLT and brighter than $B=16.5$.  At this time a second
spectrum was still lacking for 242 WDs, but observing time has been
granted to complete these observations. Currently we could check 875
stars for RV variations, and detected $\approx$100 new DDs, 16 are
double-lined systems (only 6 were known before). The great advantage
of double-lined binaries is that they provide us with a well
determined total mass (cf.\ below).  Our sample includes many short
period binaries (some examples are discussed in
Sect.~\ref{s:followup}), several with masses closer to the
Chandrasekhar limit than any system known before, including one
possible SN\,Ia progenitor candidate (cf.\ Fig.~\ref{f:pm}).  In
addition, we detected 19 RV variable systems with a cool main sequence
companion (pre-cataclysmic variables; pre-CVs). Some examples of
single-lined and double-lined DDs are shown in Fig.~\ref{f:halpha}.
Our observations have already increased the DD sample
by a factor of seven.  

Although important information like the periods, which can only be
derived from follow-up observations (see below), are presently lacking
for most of the stars, the large sample size already allows us to draw
some conclusions. (Note that fundamental WD parameters like masses are
known from spectral analysis; Koester et al.\ 2001). One interesting
aspect concerns WDs of non-DA classes. Since no sharp NLTE cores are
available, non-DA WDs were not included in most RV surveys.  SPY is
the first RV survey which performs a systematic investigation of both
classes of WDs. The use of several helium lines enables us to reach an
accuracy similar to the DA case. Our result is that the binary
frequency of the non-DA WDs is not significantly different from the
value determined for the DA population.

\paragraph{Parameters of double degenerates:}
\label{s:followup}

Follow-up observations of this sample are mandatory to exploit its
full potential. Periods and WD parameters must be determined to find
potential SN\,Ia progenitors among the candidates.  Good statistics of
a large DD sample will also set stringent constraints on the evolution
of close binaries, which will dramatically improve our understanding
of the late stages of their evolution.


The secondary of most DD systems has already cooled down to
invisibility. These DDs are single-lined spectroscopic binaries (SB1). Our
spectroscopic follow-up observations allow us to determine the orbit of the
primary component (i.e.\ the period $P$ and the RV amplitude
$K_1$). The mass of the primary $M_1$ is known from a model atmosphere analysis
(Koester et al.\ 2001). Constraints on the mass of the secondary $M_2$ 
can be derived from the mass function.
For a given inclination angle $i$ the mass of the secondary can be
computed. However, $i$ is rarely known, but the result for $i=90^\circ$ 
yields a
lower mass limit. For a statistical analysis it is useful to adopt the most
probable inclination $i = 52^\circ$. We have plotted the single-lined systems
with the resulting system mass in Fig.~\ref{f:pm}. Note that two SB1
binaries have
probably combined masses in excess of the Chandrasekhar limit. However, the
periods are rather long preventing merging within a few Hubble times.


\begin{figure}[tbh]
\plotone{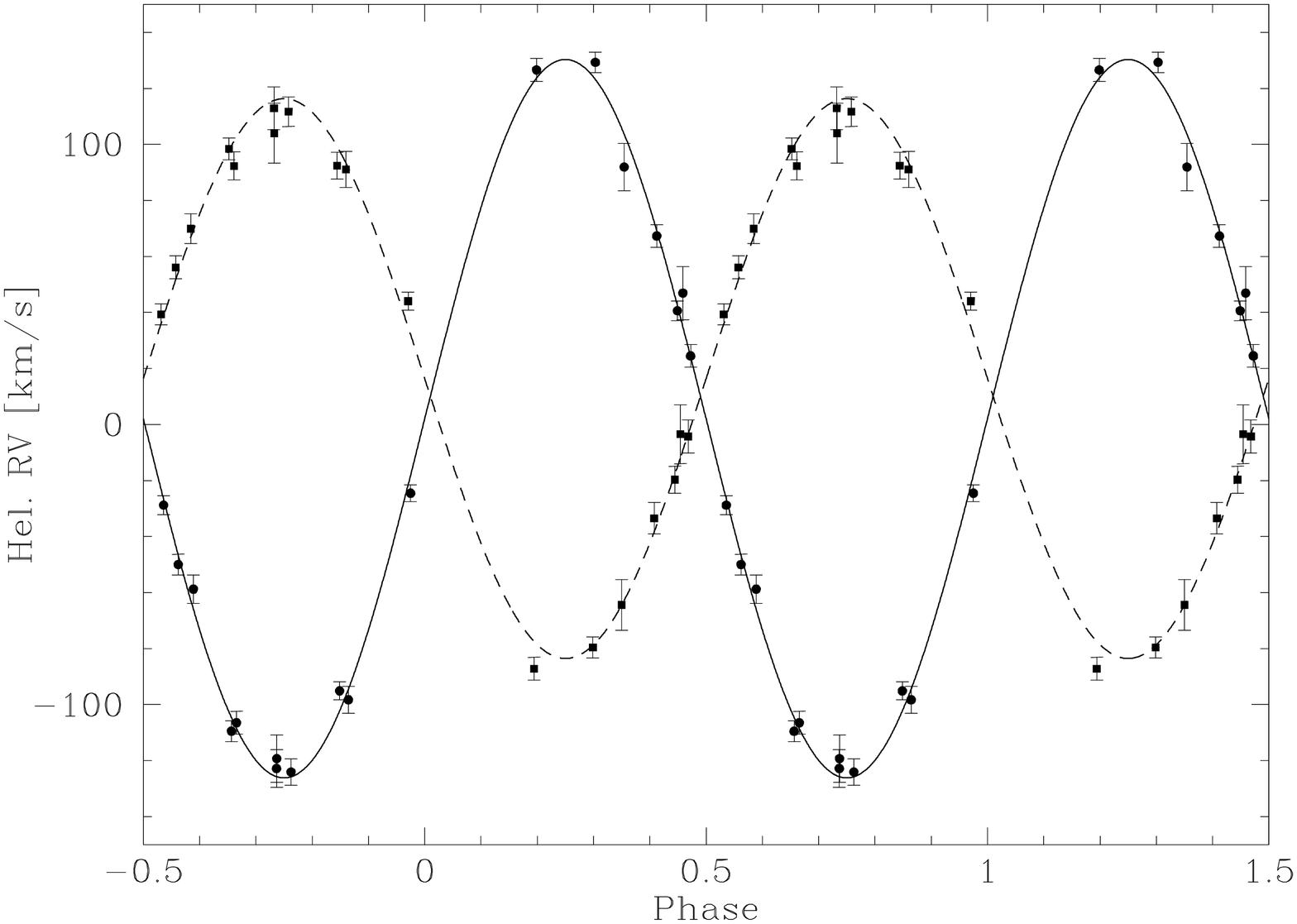}
\caption{Measured RVs as a function of orbital phase and
fitted sine curves for HE\,1414-0848. Circles and solid line/rectangles
and dashed line 
 indicate the less/more massive component 1/2. Note the difference of the
``systemic velocities'' $\gamma_0$ between both components caused by
gravitational redshift.}
\label{f:HE1414rv}
\end{figure}

Sometimes spectral features of both DD components are visible
(Fig.~\ref{f:HE1414ha}), i.e.\ these are
double-lined spectroscopic binaries (SB2). 
As an example for other double-lined systems we discuss here the 
DA+DA system HE\,1414$-$0848 (Napiwotzki et al.\ 2002).
On one hand 
the analysis is complicated for double-lined systems, 
but on the other hand the spectra contain more
information than spectra of single-lined systems. The RVs of both WDs
can be measured, and the orbits of both individual components can be determined
(Fig.~\ref{f:HE1414rv}). 
For our example HE\,1414$-$0848 we derived a period of 
$P = 12^{\mathrm{h}} 25^{\mathrm{m}} 44^{\mathrm{s}}$ and
semi-amplitudes $K_1 = 127$\,km\,s$^{-1}$ and $K_2 = 96$\,km\,s$^{-1}$.
The ratio of velocity amplitudes is directly related to the
mass ratio of both components:
${M_2}/{M_1} = {K_1}/{K_2} = 1.28\pm 0.02$.
However, additional information is needed before the absolute masses can be
determined. There exist two options to achieve this goal in double-lined DDs. 
From Fig.~\ref{f:HE1414rv} it is evident that the ``system velocities''
derived for components 1 and 2 differ by 14.3\,km/s, 
much more than naively expected from the error bars. However, this is easily
explained by the mass dependent gravitational redshift of WDs
$z = {GM}/{Rc^2}$
This offers the opportunity to determine masses of the individual WDs
in double-lined DDs. For a given mass-radius relation 
gravitational redshifts
can be computed as a function of mass. Since the mass ratio is already
known from the amplitude radio,
only one combination of masses can fulfil both constraints. In the case of
He\,1414$-$0828 we derived individual masses $M_1 = 0.55\pm 0.03 M_\odot$ 
and $M_2 = 0.71\pm 0.03M_\odot$. The sum of both WD
masses is $M=1.26\pm 0.06M_\odot$. Thus HE\,1414$-$0848 is a massive DD with a
total mass only 10\% below the Chandrasekhar limit.

This method cannot be used if the systems
consist of WDs of low mass, for which the individual gravitational
redshifts are small,  or if their masses are too similar, because
the redshift differences are very small and
this method cannot be used to determine absolute masses.  Another
method, which works in these cases as well, are model atmosphere
analyses of the spectra to determine the fundamental parameters,
effective temperature and surface gravity $g = G\,{M}/{R^2}$, of the
stars.  Because this system is double-lined the spectra are a
superposition of both individual WD spectra.  A direct
approach would be to disentangle the observed spectra by deconvolution
techniques into the spectra of the individual components.  Then we
could analyse the spectra by fitting synthetic spectra developed for
single-lined WDs to the individual line profiles.  Such
procedures were successfully applied to main sequence double-lined
binaries (as discussed elsewhere in these proceedings).  However, they
have not been tested for WDs, for which the wavelength shifts
caused by orbital motions are much smaller than the line widths of the
broad Balmer lines.  Therefore we choose a different approach for our
analysis of double-lined DD systems. We developed the programme {\sc
fitsb2}, which performs a spectral analysis of both components of
double-lined systems.  It is based on a $\chi^2$ minimisation
technique using a simplex algorithm.  The fit is performed on all
available spectra covering different spectral phases simultaneously,
i.e.  all available spectral information is combined into the
parameter determination procedure.

\begin{figure}[tb!]
\plotone{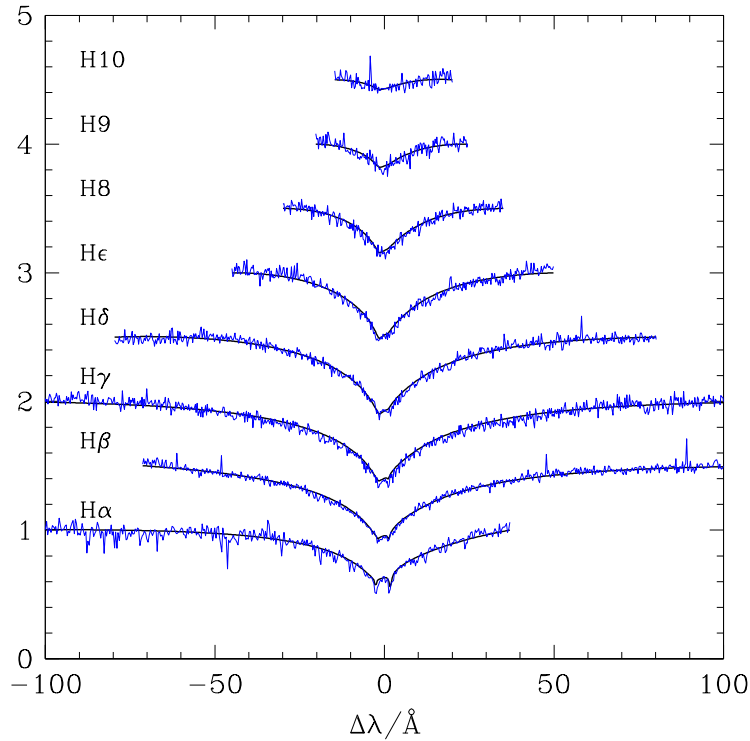}
\caption{Model atmosphere fit of the Balmer series of HE\,1414$-$0848 with
  {\sc fitsb2}. 
This is only a sample fit. All available spectra, covering different
  orbital phases, were used simultaneously.}
\label{f:fitsb2}
\end{figure}

The total number of
fit parameters (stellar and orbital) is high. Therefore we fixed as many
parameters as possible before performing the model atmosphere analysis.
We have kept the RVs of the
individual components fixed according to the RV curve. 
Since the mass ratio is already accurately determined from the RV
curve we fixed the gravity ratio.
The remaining fit parameters are the effective
temperatures of both components and the gravity 
of the primary.
The gravity of the secondary is calculated from that of the primary
and the ratios of masses and radii. While the former is known from the
analysis of the RV curve, the latter has to be estimated from
mass-radius relations.
The relative contributions of both stars is determined by their
radii and surface fluxes.
The flux ratio in the V-band is calculated from the actual
parameters and the model fluxes are scaled accordingly.
The individual contributions are updated
consistently as part of the iteration procedure.
The results for HE\,1414$-$0848 are $T_{\mathrm{eff}}$/$\log g$ =
8380\,K/7.83 and 10900\,K/8.14 for components 1 and 2. A sample fit is
shown in Fig.~\ref{f:fitsb2}. The derived $\log g$ values are in good
agreement with the values corresponding to the masses derived from the
RV curves: $\log g = 7.92$ and 8.16 respectively.
 
We have plotted HE\,1414$-$0848 as well as our other results on double-lined
systems in Fig.~\ref{f:pm}.  Note that one double-lined system is probably a
SN\,Ia progenitor. However, the RV curve of the hotter component is very
difficult to measure causing the large error bars. Observing time
with the far-UV satellite FUSE has been allocated, which will enable us to
measure more accurate RVs. More individual objects are discussed in
Napiwotzki et al.\ (2001b) and Karl et al.\ (2003).

\section{Concluding remarks}

The large programme part of SPY has now been completed with some
observations underway to complete the observations of the WDs with
only one spectrum taken during the survey.  We increased the number of
WDs checked for RV variability from 200 to 1000 and multiplied the
number of known DDs by more than a factor of six (from 18 to $\approx$120)
compared to the results achieved during the last 20 years.  Our sample
includes many short period binaries (Fig.~\ref{f:pm}),
several with masses closer to the
Chandrasekhar limit than any system known before, greatly improving the
statistics of DDs. 
We expect this survey to produce a sample of $\approx$150
DDs. 

This will allow us not only to find several of the long sought potential
SN~Ia
precursors (if they are DDs), but will also provide a census of the
final binary configurations, hence an important test for the theory of
close binary star evolution after mass and angular momentum losses
through winds and common envelope phases, which are very difficult to model.
An empirical calibration provides the 
most promising approach. A large sample of binary WDs 
covering a wide range
in parameter space is the most important ingredient for this task.

Our ongoing follow-up observations already revealed the existence of three
short period systems with masses close to the Chandrasekhar limit, which will
merge within 4\,Gyrs to two Hubble times. Even if it will finally turn out
that the mass of our most promising SN\,Ia progenitor candidate system is
slightly below the Chandrasekhar limit, our results already allow a
qualitative evaluation of the DD channel. Since the formation of a system
slightly below Chandrasekhar limit is not very different from the formation of
a system above this limit, the presence of these three systems alone provides
evidence (although not final proof)
that potential DD progenitors of SN\,Ia do exist.



\begin{references}
\reference Bragaglia A., Greggio, L., Renzini, A., \& D'Odorico, S. 1990, ApJ 
    365, L13
\reference Ergma, E., Fedorova, A.V., \& Yungelson, L.R. 2001, A\&A
    376, L9
\reference Foss, D., Wade, R.A., \& Green, R.F. 1991, ApJ 374, 281
\reference Iben, I.Jr., \& Tutukov, A.V. 1984, ApJS 54, 335
\reference Iben, I.Jr., Tutukov, A.V., \& Yungelson, L.R. 1997, ApJ 475, 291
\reference Karl, C.A., Napiwotzki, R., Nelemans, G., et al. 2003, A\&A 410,
           663
\reference Koester, D., Napiwotzki, R., Christlieb, N., et al.
        2001, A\&A 378, 556
\reference Livio, M. 2000 in ``Type Ia Supernovae: Theory and Cosmology'', 
        Cambridge Univ. Press, p.~33
\reference Marsh, T.R. 2000, NewAR 44, 119
\reference Maxted, P.F.L., Marsh, T.R. 1999, MNRAS 307, 122
\reference Maxted, P.F.L., Marsh, T.R., \& Moran, C.K.J. 2000a, MNRAS, 319, 305
\reference Maxted, P.F.l., Marsh, T.R., \& North, R.C. 2000b, MNRAS
    317, L41
\reference Napiwotzki, R., Christlieb, N., Drechsel, H., et al.
        2001a, AN 322, 401
\reference Napiwotzki, R., Edelmann, H., Heber, U., et al. 2001b, 
        A\&A 378, L17
\reference Napiwotzki, R., Koester, K., Nelemans, G., et al. 
        2002, A\&A 386, 957
\reference Nelemans, G., Yungelson, L.R., Portegies Zwart, S.F., \&
        Verbunt, F. 2001, A\&A 365, 491
\reference Robinson, E.L., \& Shafter, A.W. 1987, ApJ 322, 296
\reference Saffer, R.A., Livio, M., \& Yungelson, L.R. 1998, ApJ 502, 394
\reference Webbink, R.F. 1984, ApJ 277, 355
\reference Whelan, J., Iben, I.Jr. 1973, ApJ 186, 1007
\end{references}
\end{document}